# Ultrafast hot-electron induced quenching of Tb4f magnetic order.

T. Ferte[1], N. Bergeard[1], G. Malinowski[2], R. Abrudan[3], T. Kachel[3], K. Holldack[3], M. Hehn[2], and C. Boeglin[1]

[1]*Université de Strasbourg, CNRS, Institut de Physique et Chimie des Matériaux de Strasbourg, UMR 7504, F-67000 Strasbourg, France.*

[2]*Institut Jean Lamour, Université Henri Poincaré, Nancy, France.*

[3]*Institut für Methoden und Instrumentierung der Forschung mit Synchrotronstrahlung, Helmholtz-Zentrum Berlin für Materialien und Energie GmbH, Albert-Einstein-Str. 15, 12489 Berlin, Germany.*

**Abstract**:

We have investigated ultrafast quenching of the Tb4f magnetic order in $Co_{74}Tb_{26}$ alloys, induced by femtosecond hot-electrons pulses. The hot-electron pulses were produced in specific non-magnetic capping layers by infrared fs laser pulses. Our experimental results show that sub-picosecond dynamics of Tb4f magnetic moments can be induced by non-thermal and thermal hot-electrons. We further demonstrate that the demagnetization efficiencies of non-thermal and thermal hot electrons are similar. However, the characteristic demagnetization times show values of 0.35 ps for non-thermal hot-electrons excitations and 1.2 ps for thermal hot-electrons excitations. We explain this temporal elongation by the propagation time of thermal hot-electrons through the 15 nm thick CoTb film.

## A. Introduction:

Determining the ultimate speed for deterministic control of magnetization manipulation is a hot-topic in modern magnetism, mainly due to technological application in data manipulation [1]. In this context, the discovery in 1996 [2] of ultrafast magnetization quenching upon femtosecond laser excitation has driven intensive experimental and theoretical works [3]. In the frame of the thermodynamic 3 temperatures model (3TM), the laser pulse injects its energy in the electronic system that relaxes due to the coupling with the lattice (electron-phonon coupling) and with the spins (electrons-spin coupling) [2]. At the sub-picosecond time scale, the actual phenomenological model as proposed by Beaurepaire et al. does neither address the conservation of the angular momentum nor the microscopic mechanisms driving the dissipation of the angular momentum [2]. However, both aspects are heavily debated in many experimental and theoretical works. In 2009, Koopmans et al. proposed a microscopic 3TM, in which spin-flips resulting from the collision between laser-excited electrons with phonons and/or impurities have been introduced to account for the angular momentum conservation [4]. Even though this model accurately reproduced some trends of the demagnetization in transition metal [5, 6], rare-earth elements [4] and in many magnetic alloys [7-9], doubts have been casted upon the efficiency of spin-flip scattering in ultrafast demagnetization [10-11]. Alternatively, the transport of spin-polarized hot-electrons has been proposed to explain ultrafast demagnetization [12, 13, 14]. In metals, the pump IR pulses create a strongly out-of-equilibrium electronic distribution called the "non-thermal hot-electrons". These non-thermal hot-electrons can propagate at the Fermi velocity over tens of nanometers. Due to electron-electron collisions, the non-thermal hot-electrons reach an internal equilibrium characterized by a Fermi-Dirac distribution, called "thermal hot-electrons". Malinowski et al. [12] proposed a ballistic transport of spin-polarized hot-electrons as an efficient channel for angular momentum dissipation during the ultrafast demagnetization. Within this model the different amplitudes and time scales which were experimentally observed have been explained [12]. Recently, Battiato et al. proposed a theoretical model in which the demagnetization is entirely explained by the superdiffusive transport of spin-polarized hot-electrons [13, 14]. This model has been further sustained by experimental results [15-17]. Since then, plethora of experimental and theoretical works evidenced that local (such as spin-flip) and non-local mechanisms (transport of hot-electrons) are both important ingredients to explain the ultrafast magnetization dynamics. Today the controversy lies in the relative importance of the proposed mechanisms [18-21].

Meanwhile, first attempts towards femtosecond spin-transfer torques by ballistic hot-electrons have been made by Schellekens et al. [24] whereas Choi et al. showed that ultrafast spin manipulation can be obtained through electronic thermal transport [25, 26]. Such ultrashort hot-electron pulse induced spin manipulations opens new routes towards femtosecond spintronics [27]. These potential applications motivate the community to investigate the details of the interactions between the hot-electron pulses and the magnetic layers. Recent studies have now unambiguously confirmed the existence of hot-electron induced ultrafast demagnetization in 3d transition metals [16, 22, 23, 9]. These experiments have reported ultrafast magnetization dynamics induced either by non-thermal hot-electrons [9, 16] or thermal hot-electrons [23]. Unfortunately, it is not straight forwards to compare the demagnetization efficiencies induced by non-thermal and thermal hot-electrons from those published data since different metals were studied. Furthermore, only the cases of ultrafast demagnetization in 3d elements have been studied.

Previous experiments revealed fundamental dissimilarities in the laser induced ultrafast magnetization dynamics of transition metals (TM) [28] and rare-earth (RE) layers [29]. These differences are related to the localization and hybridization of electron orbitals carrying the spin momentum (respectively the 4f and 3d electrons) and the magnetic anisotropy [30, 29]. It is well known that the magnetic transition metals (Ni, Fe, Co) show faster demagnetization than rare-earth elements (Gd, Tb, Dy…), a tendency which is also evidenced in RE-TM alloys where this difference is even more intriguing [31, 33]. Therefore, the itinerant 3d and localized 4f induced dynamics has been intensively studied in such alloys [33-35]. However, in these alloys the hot-electron induced dynamics is still unexplored. The similarities and differences between IR laser and hot-electron induced 3d and 4f spin dynamics should give insights into the driving mechanisms in both cases.

In this work, we provide a study of the hot-electron induced dynamics of the 4f magnetic moments in a $Co_{74}Tb_{26}$ alloy. For this, we used the element sensitivity of time-resolved X-Ray Magnetic Circular Dichroism (XMCD) [32, 33]. Furthermore, comparing the ultrafast demagnetization measured for different samples, we distinguish the specificities of the demagnetization which are induced by non-thermal or by thermal hot-electrons. Among other RE-TM alloys, the $Co_{74}Tb_{26}$ alloy is well suited for our study because it shows a fast IR laser induced 4f demagnetization dynamics with a short characteristic time of $\tau = 0.28$ ps [32, 33]. This fast demagnetization dynamic enhances our sensitivity to other (or additional) excitation sources for which slower dynamics are induced.

**B. Static and time-resolved experiments**:

The 15 nm thick $Co_{74}Tb_{26}$ (label CoTb in the text for commodity) alloys were deposited by DC-magnetron sputtering on Si3N4 membranes of 200 nm thickness and capped with 3 different metallic layers: Pt(2)/Cu(3.5) (sample 1- Figure 1), Pt(4)/Cu(70) (sample 2- Figure1) and Pt(4)/Cu(50)/Pt(10)/Cu(3.5) (sample 3- Figure1) (units are in nm). In samples 2 and 3, the IR femtosecond laser pulses are absorbed in the Pt(4)/Cu(X) capping layers generating non-thermal hot-electron pulses, as described by Bergeard et al [9]. According to our calculations, based on the variation of the Poynting vector, the CoTb layers absorb 16.7%, 0.11% and 0.36% of the incident IR intensity for the samples 1, 2 and 3 respectively (method in the supplements of [9]). Thus, the choice of the different capping layers ensures negligible direct IR excitations in the CoTb alloys for samples 2 and 3 [9]. The thickness of the capping layers is nevertheless limited by the X-ray transmission at the Tb $M_5$ absorption edges which is 15% in our experiment. In Cu, the thermalization time of hot-electrons is above 500 fs [36]. Therefore, in sample 2, the photo-excited non-thermal hot-electrons can propagate at the Cu Fermi velocity towards the $Co_{74}Tb_{26}$ before any thermalization of the hot-electrons [37, 38]. In sample 3, the additional Pt(10) layer acts as a filter for the non-thermal ballistic hot-electrons which are photo-excited in the upper Pt(4)/Cu(50) capping layers. This is confirmed by the electron attenuation length of ~5 nm [39]. As previously demonstrated by Choi et al, the photo-excited hot-electron pulses generated in the Pt(4nm)/Cu(50nm) drive the electron distribution of the Pt(10) layer in a strongly out-of-equilibrium distribution as efficiently as IR fs laser pulses [40]. The hot-electrons are rapidly thermalized since the electron thermalization times in Pt is much shorter (<200 fs) than those expected in noble metals (> 500 fs) [41]. The energy absorbed in the Pt(10) layer is then transferred to the adjacent layers through heat current mediated by diffusive thermal hot-electrons [25]. As a consequence, direct laser excitations are investigated in sample 1 whereas the excitations induced by non-thermal and thermal hot-electron are investigated by comparing the dynamics in samples 2 and 3. To guarantee similar magnetic properties (i.e. anisotropy, magnetic moments) for all samples, we grew a 3.5 nm Cu film on top of the CoTb alloys in samples 1 and 3. Static XMCD spectroscopy has been performed using ALICE spectrometer at the PM3 beamline at the BESSY II synchrotron radiation source operated by the Helmholtz-Zentrum Berlin [42]. The X-ray Absorption Spectra (XAS) were measured by monitoring the transmitted X-ray intensities as a function of the photon energy. The X-ray beam and the applied magnetic field of ±0.7 T were aligned along the sample normal. Two XAS spectra have been recorded with opposite film magnetizations revealing the XMCD at the $TbM_5$ edge (Figure 2a). Hysteresis loops were recorded by tuning the X-ray energy to the $TbM_5$ edge (Figure 2b). These hysteresis evidence square loops and coercive fields of $H_C$=0.33 T well below the maximum field of the experimental set-up at the FEMTOSPEX end-station [43]. Finally, the XMCD measurements show that all samples display almost similar static magnetic properties at the $TbM_5$ edge (out of plane magnetic anisotropy, square loops, coercive fields, and the values of the magnetic moments) (Fig 2a,b).

The time-resolved XMCD experiments were performed at the femtoslicing beam line of the BESSY II synchrotron radiation source at the Helmholtz-Zentrum Berlin [43]. The experiments were performed in transmission, under an applied magnetic field of ±0.55T, using the same geometry as in the static experiments. The experiment requires a pump-probe setup working at 3 kHz, where femtosecond IR laser pulses (3 kHz) were used as the pump while the circularly polarized X-ray pulses (6 kHz) were used as the probe. This procedure allows for probing both, the pumped and unpumped transient states. The duration of IR laser and X-ray pulses in the femtoslicing operation mode were 60 fs and 100 fs respectively, which ensures a global time resolution of ~130 fs [28, 30]. The magnetization dynamics have been recorded by monitoring the transmission of the circularly polarized X rays tuned to the core level absorption edge $TbM_5$ as a function of the pump-probe delay. The multi-sample holder of the experimental chamber allows measuring alternatively from samples 1 (direct pumping by IR pulses) to sample 2 and 3 (indirect excitation). The laser beam diameter of ~500 μm ensures a homogeneous excitation over the X-ray probed area (~150 μm). The laser power was P= 8 mW for sample 1 and P= 24 mW for sample 2 and sample 3. The cryostat temperature was set to 250K in order to compensate the laser DC-heating, which ensures a base temperature of T ~300K at negative delays, well below the temperature of magnetic compensation of the CoTb alloy films ($T_{comp}$ ~550K [32]).

**C. Time-resolved experiments**:

The transient XMCD data recorded at the TbM$_5$ edge are shown in Figure 3 for the samples 1, 2 and 3. The ultrafast quenching of the Tb 4f magnetic order is observed for all samples. The data are normalized by the XMCD at negative delay and adjusted by classical exponential fits convoluted by a Gaussian function which accounts for the experimental time resolution [30, 33]. We have characterized the dynamics in the 3 samples by the following parameters: the demagnetization amplitudes q, the onset of the ultrafast demagnetization t and the characteristic demagnetization time τ. These parameters, and their error bars, are extracted from the fits and summarized in table 1.

For sample 1, the Tb demagnetization is driven by the direct IR laser excitation and the ultrafast dynamics is used during our experiment as a reference for t0. Since the pump conditions have been adjusted to reach almost the same demagnetization amplitude (q~0.70) as in our previous work, we will use the characteristic demagnetization time τ=0.28±0.03 ps as the reference value for direct IR laser induced demagnetization. We note that the quenching of the magnetic order is reached within the first ps after the laser excitation (empty circles in figure 3) [32, 33].

For samples 2 and 3, the observed ultrafast Tb demagnetization is not driven by the IR laser excitation. The demagnetization amplitudes and the onset of demagnetization are almost similar for both samples but the characteristic demagnetization time τ is much larger for sample 3 τ =1.2±0.4 ps than for sample 2 τ =0.35±0.1 ps (table 1). This difference raises intriguing questions about the origin of the elongation observed when the 10 nm thick Pt layer is inserted in the thick Cu layer, on top of CoTb. The IR laser power used for the indirect excitations (24 mW) is much larger than the one used for the direct excitation (8mW). However, the amplitudes of the demagnetization in the CoTb films are only of q~52% and q~36% for sample 2 and samples 3 respectively.

**D. Discussion**:

It has been shown in the literature that for 3d metals the direct laser excitation is not the only excitation process which is able to induce sub-picosecond magnetization dynamics [16, 22, 9, 23]. In this work, we extend these conclusions to the case of localized 4f magnetic moments in RE-TM alloys. We have used the same capping layers on our RE-TM films as Bergeard et al. on their CoPt multilayers [9]. In their work, the onset of the demagnetization is shown to be delayed in the buried magnetic multilayer, in respect with the IR excitation by ~120 fs whereas the characteristic demagnetization time τ is longer by ~ 30 fs for a 70 nm Cu capping thickness. The observed elongation of the characteristic demagnetization time τ originates mainly from the spatial elongation of the ballistic transport of non-thermal hot-electron pulses during the propagation towards the FM multilayer [9]. Following the description of demagnetization induced by ballistic transport of hot-electrons provided by Bergeard et al, the demagnetization onset of the Tb sublattice in sample 2 should be ~120 fs. This onset of 120 fs is consistent with our experimental value of 250 fs considering our large error bars (±200 fs).

The demagnetization amplitude in sample 2 is q= 52% at the TbM$_5$ edge, which is surprisingly large considering the large Co$_{74}$Tb$_{26}$ film thickness (15 nm) and the limited inelastic mean-free path (IMFP) of the incoming ballistic hot-electrons ($\lambda_{up}$=6.5 nm for majority spins and $\lambda_{down}$=1.2 nm for the minority spins in Co [39]). We have adapted equation (1) from [44], to estimate the in-depth profile of the energy deposited by the incoming non-thermal hot-electron in the CoTb layer. We have injected the spin-dependent IMFP instead of IR penetration depth in the formula. We estimate that ~80% of the energy carried by hot-electrons with minority spin character is deposited within the first 4 nm of the Co$_{74}$Tb$_{26}$ layer. We also estimate that less than 10% of the energy carried by hot-electrons with majority spin character is deposited at the bottom of the alloys. The energy injected by ballistic non-thermal hot-electrons is transported away from the "surface" towards the "bulk" of the Co$_{74}$Tb$_{26}$ alloy by diffusive non-thermal hot-electrons at the Fermi velocity, as shown by Tas et al. [44]. Ultrafast demagnetizations induced by non-thermal hot-electrons (in the ballistic and diffusive regimes) have been reported previously in 3d transition metals [9, 16]. Therefore, extending this description to our sample 2, we can propose two contributions to the ultrafast demagnetization measured in the Tb 4f sublattice. The first is described by inelastic scattering of the incoming ballistic non-thermal hot-electrons with the electrons at the Fermi level of the Co$_{74}$Tb$_{26}$ alloy. The second is described by the multiple scattering events experienced by the non-thermal hot-electrons in the Co$_{74}$Tb$_{26}$ film. We show that the scattering events experienced by the non-thermal hot-electrons drive the ultrafast demagnetization in sample 2. Similar to laser induced ultrafast dynamics in RE layers [29, 45] and RE-TM alloys [31, 32, 33], the hot-electrons excite the 3d and 5d electrons in the Co$_{74}$Tb$_{26}$ alloy. The scattering events experienced by the hot-electrons drive the magnetization dynamics either by producing spin-polarized superdiffusive hot-electrons as claimed by Eschenlohr et al. [16] or by inducing spin-flip as claimed by Bergeard et al [9]. The demagnetization of the localized 4f moments is mediated via the RKKY indirect exchange coupling [29, 46]. However, the description of the microscopic mechanisms which explains the observed demagnetization is beyond the scope of this article.

Concerning the value of the characteristic demagnetization time $\tau$ = 350±100 fs extracted from sample 2, we note that it is close to the value measured for direct IR excitations in Co$_{74}$Tb$_{26}$ (sample 1, $\tau$=280 ± 30fs). According to Bergeard et al., the elongation of the ballistic non-thermal hot-electron pulses in our capping layer, compared to the IR pulse, does not exceed 30 fs [9]. The velocity of non-thermal hot-electrons (ballistic and diffusive) in Co is ~0.25 nm/fs [47, 49]. Therefore, the propagation time of the non-thermal hot-electrons in the 15 nm Co$_{74}$Tb$_{26}$ layer is ~60 fs (15 / 0.25 = 60 fs) which is well below the characteristic electronic thermalization time of ~100 fs in Co [48]. The non-thermal hot-electron propagation time of 60 fs demonstrate that the different atomic layers in our Co$_{74}$Tb$_{26}$ film are not simultaneously excited. The difference with the instantaneous excitation induced by IR lasers, is due to a notably lower electronic group velocity (~0.25 nm/fs) compared to the speed of light. Adding up the individual contributions we found a global elongation of the characteristic demagnetization time $\Delta\tau$ = 90 fs (stretching in capping layer 30 fs + propagation time in the Co$_{74}$Tb$_{26}$ alloy 60 fs). Thus, considering $\tau$~280 fs for direct IR excitation and $\Delta\tau$ = 90 fs, we calculate a global demagnetization time of $\tau$~370 fs. We conclude that our value

τ=350±100 fs measured in sample 2 can be justified by the transport of non-thermal hot-electrons towards and within the $Co_{74}Tb_{26}$ film. Finally, we conclude that in sample 2, at the TbM5 edge, the ultrafast demagnetization which is characterized by the demagnetization onset, amplitude and characteristic time is consistent with an ultrafast dynamic induced by non-thermal hot-electrons.

In order to confirm the possibility to induce ultrafast demagnetization in the Tb 4f sublattice by using thermal hot-electrons as proposed by Choi et al. [40] we performed the pump-probe experiment with sample 3. To compare the efficiency of demagnetization induced by non-thermal or thermal hot-electrons we have performed the measurements using the same $Co_{74}Tb_{26}$ alloy but with a specifically designed capping (sample 3). In sample 3, a Pt layer of 10 nm thickness has been inserted between the Pt(4nm)/Cu(50nm) and the CoTb alloy layers (figure 1) in order to produce a heat current of thermal hot-electrons [25]. A demagnetization q ~36% is observed for the same laser power we used for sample 2. Following previous works published by Choi et al. [25], we can attribute the demagnetization of sample 3 to the energy transfer from the excited Pt(10) layer towards the CoTb alloy, mediated by diffusive thermal hot-electrons. However, a quantitative estimation of the energies injected into the CoTb alloy for sample 2 and 3 is difficult. The energy transported by the hot-electrons dependent on the material and on the interfaces present along the trajectory of the electrons [40]. The presence of additional interfaces in sample 3 (Cu/Pt, Pt/Cu and Cu/CoTb) may however qualitatively explain the reduction of the demagnetization amplitude compared to sample 2 (Cu/CoTb) (sample 3: q~36% and sample 2: q~52%).

According to Hohlfeld et al, the order of magnitude for the velocity of the diffusive thermal hot-electrons in metal is ~0.01 nm/fs [49]. Thus, the propagation time of these electrons in the CoTb 15nm is ~1.5 ps in qualitative agreements with our experimental measurements. Therefore, the longer characteristic demagnetization time in sample 3 (τ = 1.2±0.4 ps) compared to sample 2 (τ = 0.35±0.1 ps) is qualitatively explained by the slower propagation of thermal diffusive hot-electrons. We notice the similarity of our results with the work of Vodungbo et al. who have studied the laser induced demagnetization in Al(40)/CoPd [22] evidencing a 3 times longer characteristic time τ compared to IR excitations. Comparing these results with the work of Salvatella et al. suggests that the ultrafast demagnetization of the 3d Co in the CoPd film was initiated by thermal hot-electrons [23].

The demagnetization onset measured in sample 3 is at 300±200 fs which is similar, within the error bars, to the onset measured in sample 2. By considering that the demagnetization in sample 3 is caused by the transport of thermal hot-electrons from the Pt(10) layer towards the CoTb layer, we can estimate an onset at ~320 fs which is consistent with our measured value 300±200 fs. In this calculation, we have considered the following contributions: (i) transport of the photo-excited hot-electron in Cu(50) towards the Pt(10) layer (~ + 85 fs) [9] (ii) thermalization time of hot-electron in Pt(10) (~ +200 fs [41]), (iii) heat transport through the Cu(3.5) layer (+35 fs). We use v=0.1 nm/fs for the velocity of the diffusive hot-electrons in Cu(3.5) as experimentally measured by Choi et al [25]. Different velocities in Cu (0.1 nm/fs) and Co (0.01 nm/fs) [49] are justified by the larger effective mass of the electrons in Co (~7.5 [50]) compared to Cu (~1 [51]).

We point out that Choi et al. have used thicker Pt layers (30 nm) than in sample 3, to avoid contributions of non-thermal hot-electrons upon IR excitation [25]. Therefore, we cannot assert that the thermalization of the hot-electrons is completed within our 10 nm Pt layer. Thus, we cannot exclude a contribution of residual non-thermal hot-electrons to the demagnetization of CoTb in sample 3. However, since in sample 3 we do not observe a fast characteristic demagnetization time close to the value evidenced in sample 2 ($\tau$ ~350 fs), we can confirm that the contribution of non-thermal hot-electrons is very limited. The demagnetization onset induced by such residual non-thermal hot-electrons can be estimated to ~ 113 fs (50nm Cu + 3.5 nm Cu at 0.7 nm/fs + 10 nm Pt at 0.27 nm/fs) [25, 52]. According to our large error bars of ±200 fs for the experimental onset of sample 3 (300 fs), the contribution of non-thermal hot-electrons to the onset cannot be identified.

In summary, considering the results from sample 3, we clearly evidence an efficient demagnetization by thermal hot-electrons as expected from Choi et al [40]. The signature obtained through the demagnetization characteristic times $\tau$ allows discriminating from the non-thermal hot-electrons contribution. Finally, considering our results from sample 2 and sample 3, we show that the quenching of 4f magnetic order in CoTb layers can be induced with comparable efficiencies, by non-thermal or thermal hot-electrons. The difference in the demagnetization time for sample 2 (~0.35 ps) and sample 3 (~1.2 ps) is qualitatively attributed to the velocity of the non-thermal and the thermal hot-electrons in the CoTb layers (~1 nm/fs for non-thermal hot-electrons and ~0.01 nm/fs for diffusive thermal hot-electrons) [49]. It is worth noting that the speed of sound in Pt is about 2.7 nm/ps. Therefore, we ruled out acoustic phonons triggered by hot-electrons [53] as a mechanism for the indirect ultrafast demagnetization in sample 3.

**E. Conclusions**.

We have reported on hot-electron induced ultrafast sizable demagnetization of 4f magnetic order in CoTb alloys. We provide a qualitative interpretation of our experimental findings based on the transport regime of the hot-electrons: either ballistic or diffusive. We show that the demagnetization in the CoTb layer is induced though the transport of non-thermal hot-electrons when Pt/ Cu layers are grown on the CoTb film. We further show that a thin 10 nm Pt layer is able to change the ultrafast demagnetization dynamic. The observed dynamic is compatible with thermal hot-electron mediated diffusive heat transport. Qualitative arguments allow us to reproduce the characteristic demagnetization times of $\tau$~ 0.35 ± 0.10 ps and $\tau$~ 1.2 ± 0.4 ps for both samples.

**Acknowledgements:**

We are indebted to the scientific and technical support at the Slicing facility at the BESSY II storage ring given by N. Pontius, Ch. Schüßler-Langeheine, D. Schick and R. Mitzner. The ALICE project was supported by the BMBF Contract No. 05K10PC2. The authors are grateful for financial support received from the following agencies: The French "Agence

National de la Recherche" via the projet ANR-11-LABX-0058_NIE and the project EQUIPEX UNION: No. ANR-10-EQPX-52, the CNRS-PICS program, the EU Contract Integrated Infrastructure Initiative I3 in FP6 Project No. R II 3CT-2004-50600008. Experiments were carried out on IJL Project TUBE-Davms equipments funded by FEDER (EU), PIA (Programme Investissemnet d'Avenir), Region Grand Est, Metropole Grand Nancy and ICEEL.

**Figures:**

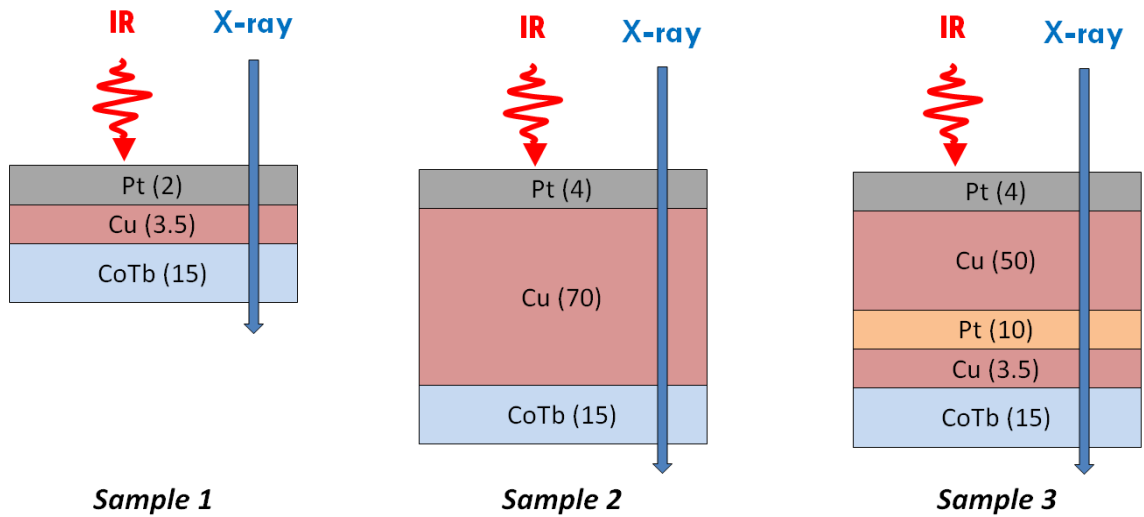

*Sample 1*  *Sample 2*  *Sample 3*

**Figure 1:** (a) Sample 1 configuration used as the reference sample and pumped with the IR laser (b) Sample configurations for sample 2 used to excite the CoTb film with ballistic hot-electrons, (c) Sample configurations for sample 3 used to excite the CoTb film with diffusive electrons.

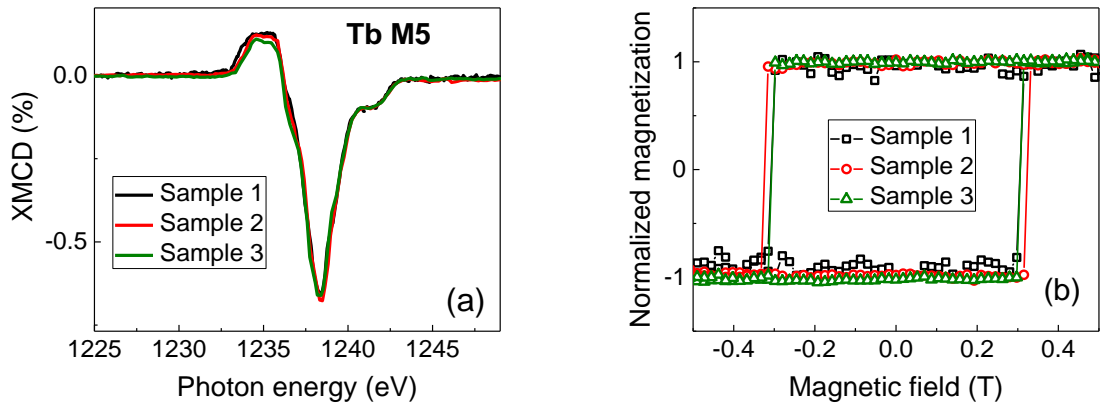

**Figure 2:** (a) XMCD spectra at the Tb $M_5$ edge and (b) hysteresis loops for the samples 1, 2 and 3. The measurements were carried out at T=300K for a normal incident of the X-ray beam in respect with the sample's surface. XMCD spectra are defined as the difference of XAS spectra obtained for two magnetic field helicities (H=0.7T). For all samples, the spectra are normalized such as the amplitude of the isotropic XAS spectrum at the $M_5$ resonance ($E_{hv}$=1237 eV) is equal to 1 (the base line at $E_{hv}$=1220eV is equal to 0). Hysteresis loops were acquired by monitoring the X-ray transmission at the Tb $M_5$ resonance ($E_{hv}$=1237 eV) as a function of the magnetic field.

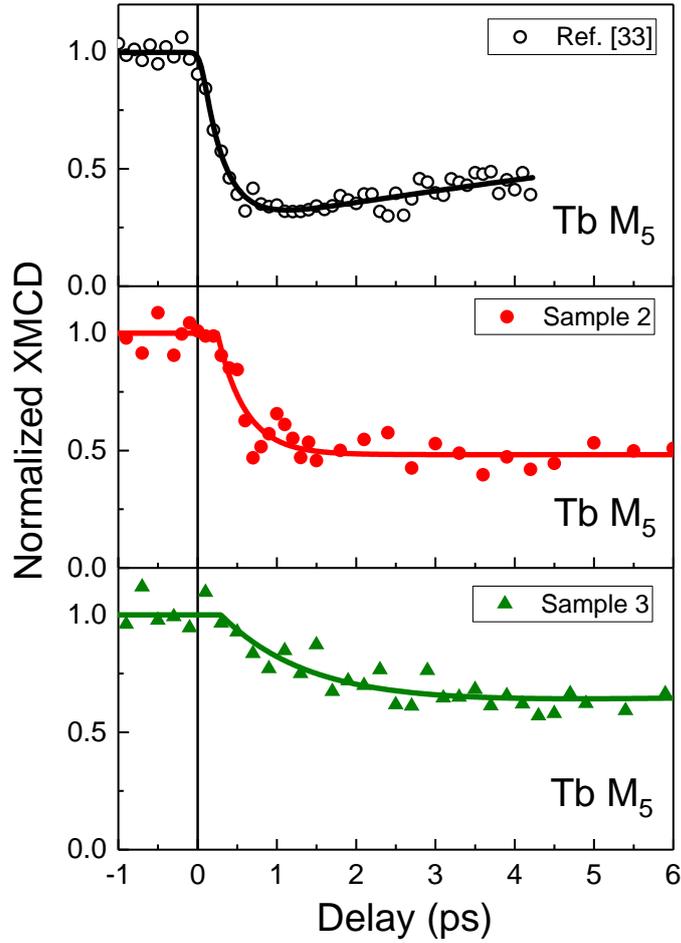

**Figure 3:** Transient normalized XMCD at the Tb $M_5$ edge for sample 1, 2 and 3. For sample 2 and 3, the solid lines are fits in the framework of the 3 temperature model. For sample 1, which is used to set the temporal overlap in this experiment, we superpose the data (open circles) and the fit of our previous Tb $M_5$ dynamics taken in the same experimental configuration [33]. The different sample configurations (sample 1, 2, and 3) show the resp. characteristic demagnetization times of τ = 0.28±0.03 ps, 0.35±0.1 ps and 1.2±0.4 ps.

Tables:

| Sample | Demagnetization amplitude q (%) | t (ps) | Characteristic demagnetization time τ (ps) |
|---|---|---|---|
| Sample 1 | 70 ± 5* | 0 ± 0.2 | 0.28 ± 0.03* |
| Sample 2 | 52 ± 8 | 0.25 ± 0.2 | 0.35 ± 0.10 |
| Sample 3 | 36 ± 6 | 0.3 ± 0.2 | 1.2 ± 0.4 |

*Table 1:* Parameters extracted from the fit functions for the 3 samples for Tb.

\* *The demagnetization amplitude and the characteristic demagnetization time are extracted from Bergeard et al. [33].*